\documentclass[10pt, conference]{FMFP2022}
\usepackage[top=1.50cm, bottom=1.50cm, left=1.884cm, right=1.884cm, includehead, includefoot, heightrounded]{geometry}
\usepackage{fancyhdr} 
\usepackage{graphicx} 
\usepackage[textfont=bf, font = normalsize]{caption} 
\usepackage{floatrow} 
\usepackage{amssymb}
\usepackage{amsfonts}
\usepackage{amsmath}
\floatstyle{plaintop} 
\restylefloat{table} 
\usepackage{caption}
\usepackage{subcaption}
\renewcommand{\headrulewidth}{0pt}

\begin{document}

\title{\LARGE{ \bf Improvement in sealing effectiveness of air curtains using positive buoyancy}}
\author{\textbf{Tanmay Agrawal, Narsing K. Jha and Vamsi K. Chalamalla}\\\\
\small{Department of Applied Mechanics, IIT Delhi, Hauz Khas, 110016, India}}

\maketitle
\thispagestyle{fancy} 
\pagestyle{plain} 

\noindent \textbf{ABSTRACT}\\
\noindent Air curtains are commonly employed in building applications to facilitate aerodynamic sealing against the exchange flow that occurs through an open doorway due to the density differences owing to buoyancy. Such situations often prevail due to temperature gradients across a doorway of an air-conditioned building, e.g., during the summer season in an Indian subcontinental situation. In the present study, we numerically investigate the performance of `positively buoyant' air curtains. In such installations,  the density of the jet fluid is larger than the density of the fluid contained within the building space. Using the two-dimensional Reynolds-averaged Navier-Stokes (2D RANS) formulation, we compute the temperature distribution in the flow domain and estimate the associated sealing effectiveness for various values of positive jet buoyancy and operating velocities of the air curtain. These estimates of sealing effectiveness are compared with that of a neutrally buoyant air curtain to assess the influence of positive buoyancy. We report an increase in sealing effectiveness of up to 10\%, whereas its peak value improves by about 5\%.\\

\noindent \textbf{Keywords:} Building flows, stratified flow, air curtains, positively buoyant jets.\\

\section{\textbf{INTRODUCTION}}\label{sec:Introduction}
\vspace{0.25cm}
\noindent The prevalence of buoyancy-induced density gradients is quite common in building flow applications, e.g., consider a doorway across which the fluids are maintained at different temperature levels. As a result, a driving force is manifested, which leads to bulk fluid transport across the door. To counter such exchange flows, air curtains (AC) are generally installed in the vicinity of a doorway to facilitate aerodynamic sealing. In addition to this inhibition, they also allow free passage for the building occupants without imposing a physical obstruction, thereby making them usable in a multitude of situations. Their other common applications include shopping complexes, food plazas, refrigerated cabinets, medical facilities, restriction of smoke in tunnel fires, etc. A typical AC installation utilizes a planar (rectangular) nozzle in which the air supply is sent by a blower after necessary flow conditioning. The supplied fluid can be taken either from within the room/building under consideration or from its ambient. Most usually, these installations are downward blowing, and in such cases, the air curtain impinges vertically on the floor. In other installations, the AC jet could be upward or sideways blowing with some angular tilt from the doorway. A brief comparison of various such arrangements was explored numerically by \textit{Gonçalves et al.} \cite{Goncalves2012} around a decade ago.\\

\noindent While the initial patent for the air curtain technology dates back to the early 20$^{\text{th}}$ century, it was only in the late 1960s that the performance of AC was first quantified systematically by \textit{Hayes and Stoecker} \cite{Hayes1969Design}. They measured velocity and temperature profiles along the air curtain axis in a full-scale experimental setup. Based on these measurements, they defined a dimensionless parameter, referred to as the deflection modulus, $D_m$, representing the ratio of the initial momentum flux contained within the AC jet and the `stack-effect' induced transverse force as a result of buoyancy difference. Mathematically, $D_m$ is written as
\begin{equation}
    D_m = \frac{\rho_0 b_0 {u_0}^2}{g H^2 (\rho_d - \rho_l)} = \frac{b_0 {u_0}^2}{g H^2 (\frac{T_0}{T_d} - \frac{T_0}{T_l})}
\end{equation}
\noindent where the subscripts $0$, $l$, and $d$ are associated with the curtain, light, and dense fluids, respectively. $T$ denotes the fluid temperature, and its density, $\rho$, has been obtained using the ideal gas law. The rectangular nozzle has a width, $b_0$, and the mean exit velocity is represented as $u_0$, assuming a top-hat distribution. Lastly, $H$ denotes the doorway height, and $g$ refers to the gravitational acceleration. Hayes and Stoecker demonstrated that the stability of an AC installation could be assessed based on the `reach' of the planar jet, which can be characterized through $D_m$. If the jet fluid impinges on the other side of the doorway, i.e., the floor of the room when the AC is downwards blowing, it is said to be stable and is deemed unstable (or breakthrough) otherwise. This transition from an unstable installation to a stable AC occurs at a minimum deflection modulus, $D_{m, min}$, the value of which is affected by the geometrical aspects ($b_0$, $H$) as well as the operating conditions ($\rho_0$, $\rho_c$, $\rho_d$). Other pathways of fluid leakage, e.g., windows for natural ventilation, etc., can also affect this critical value of $D_m$.\\

\noindent In the presence of a stack effect where $\rho_l \neq \rho_d$, the performance of an air curtain is measured by estimating the sealing effectiveness, $E$. Physically, $E$ represents the exchange flow suppression obtained after employing an air curtain in comparison to that of an open door scenario, i.e.,
\begin{equation}
    E = 1 - \frac{q}{q_{OD}}
    \label{eq:effectiveness}
\end{equation}
\noindent Here, $q$ is the volumetric exchange flow rate through the door with the air curtain installed, whereas $q_{OD}$ denotes the open-door-based exchange flow rate. Thus, the effectiveness of unity corresponds to a perfectly sealed doorway where the infiltration of the ambient fluid has been completely brought down. On the other hand, a value of zero suggests an absence of an air curtain, i.e., an open door situation. The existing literature suggests that AC can inhibit up to approximately 80\% of the total buoyancy-driven exchange flow across a doorway \cite{Goncalves2012,Costa2006,Foster2007,Frank2014,Jha2021b,Jha2020b,agrawal2021towards} based on eq. \ref{eq:effectiveness}.\\

\noindent In a typical installation, the air curtain feeds on the fluid in its vicinity, i.e., the blower sucks in the indoor fluid when the AC is installed inside the building space and vice versa. Most usually, this supply fluid does not undergo any thermal treatment before being sent to the nozzle exit and is thus neutrally buoyant for the surrounding fluid. In the present study, we numerically investigate the relevance of positive buoyancy in the context of air curtain flows, i.e., the density of the air curtain fluid is larger than that of the surrounding fluid. From a thermodynamic perspective, we consider the case when the supply fluid is cooled before blowing vertically downwards through the AC nozzle. This situation was first studied experimentally by \textit{Howell and Shibata} \cite{howell1980optimum}, who also observed a minimum deflection modulus below which the air curtain was not effective towards aerodynamic sealing. The dynamical aspects of a `negatively' buoyant air curtain were recently pursued by \textit{Frank and Linden} \cite{frank2015effects} where the air curtain was lighter than the ambient fluids. To the best of the authors' knowledge, there hasn't been any study investigating the performance of positively buoyant AC using numerical methods. Therefore, this work primarily focuses on estimating the sealing effectiveness of such air curtain installations using two-dimensional (2D) Reynolds-averaged Navier-Stokes (RANS) simulations. We seek to understand the influence of operating conditions (through varying $D_m$) on the effectiveness of these positively buoyant AC.\\

\noindent The computational setup and methodology are discussed briefly in section \ref{sec:methodology}. The results obtained from the present simulations are presented in section \ref{sec:results}. We first present the temperature distribution in the computational domain for different operating conditions, followed by the estimates of sealing effectiveness. Lastly, we present estimates of jet spillage in the building space in the presence of a transverse buoyancy. Further considerations of fluid mixing and flow structures are a work in progress and are not included in the current paper.

\section{\textbf{METHODOLOGY}}\label{sec:methodology}
\vspace{0.25cm}
\begin{figure*}[t]
\centering
\includegraphics[width=0.95\textwidth]{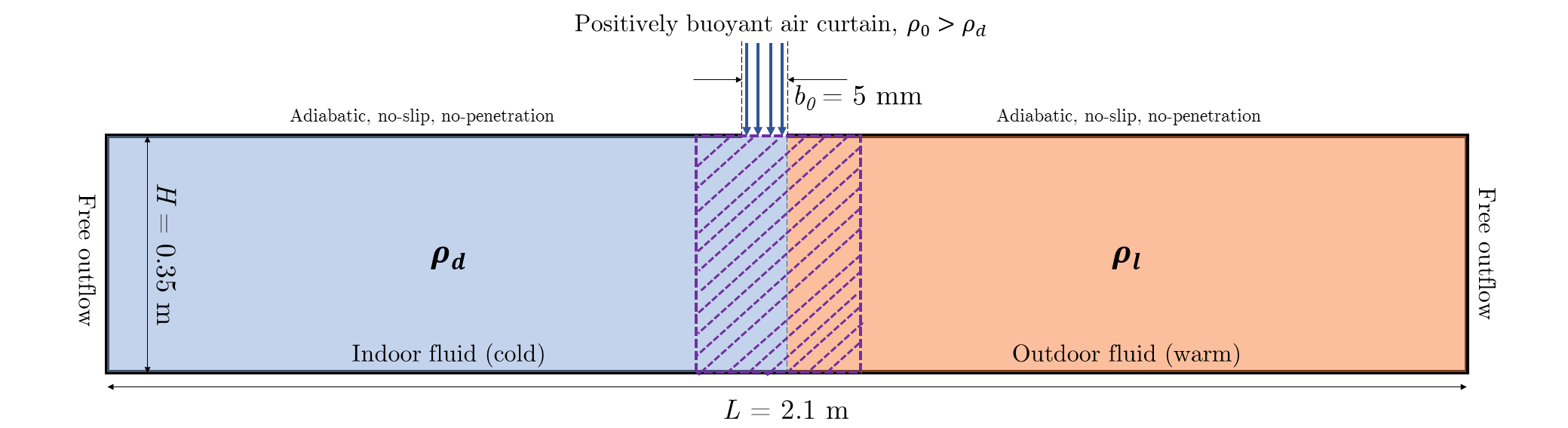}
\caption{Schematic of the simulated domain and the corresponding boundary conditions (not per scale).}
\label{fig:domain}
\end{figure*} 
\noindent Reynolds-averaged Navier-Stokes (RANS) formulation has been adopted to numerically simulate the positively buoyant air curtains. Herein, any instantaneous quantity is decomposed in its mean value and a fluctuating part. In all the simulations reported, water is chosen as the working fluid to facilitate a direct comparison with the experimental data (described later) obtained for the neutrally buoyant AC. In practical situations, different densities required in the flow can be achieved using dissolved salts (ocean), thermal gradients (atmosphere), suspended particles, etc. For the present simulations, we employ temperature inhomogeneities to create the required density difference, and the corresponding temperature levels are chosen to reflect the thermal comfort requirement in the Indian subcontinental situation. The fluid is assumed to be incompressible, and the density differences are small such that the Boussinesq approximation can be adopted. The following equations are numerically solved to simulate positively buoyant AC:
\begin{equation}
\frac{\partial{\overline{u_i}}}{\partial{x_i}}=0,
\end{equation}
\begin{equation}
\rho\frac{\partial{\overline{u_i}}}{\partial{t}}+\rho\overline{u_k}\frac{\partial{ \overline{u_i}}}{\partial{x_k}}=-\frac{\partial{\overline{p}}}{\partial{x_i}}+\frac{\partial}{\partial{x}_k}\left[\mu\frac{\partial{\overline{u_i}}}{\partial{x}_k}-\overline{\rho{u_i^\prime}{u_k^\prime}}\right],
\end{equation}
\begin{equation}
\frac{\partial{\overline{T}}}{\partial{t}}+\overline{u_k}\frac{\partial{ \overline{T}}}{\partial{x_k}}=\frac{\partial}{\partial{x}_k}\left[\kappa\frac{\partial{\overline{T}}}{\partial{x}_k}-\overline{{u_i^\prime}{T^\prime}}\right],
\end{equation}
These are mathematical representations of the conservation of mass, momentum, and energy, respectively. Here, $u_i$, $p$ and $T$ represent the $i^{th}$ velocity component ($i$ = 1, 2), pressure and the temperature respectively. The symbols $x_i$ and $t$ denote the $i^{th}$ spatial coordinate and time. $\mu$ represents the dynamic viscosity of the fluid, $\rho$, its density, and $\kappa$ denotes the thermal diffusivity. In these equations, an overbar represents an averaged (or mean) quantity, while a super-scripted $'$ represents the corresponding fluctuating quantity. The unclosed terms in these equations e.g. $-\overline{\rho{u_i^\prime}{u_k^\prime}}$, $\overline{{u_i^\prime}{T^\prime}}$, are modelled using the renormalization group (RNG) k$-\epsilon$ turbulence model \cite{yakhot1992development}. We use the RNG k$-\epsilon$ model over other turbulence closure models based on its superiority to simulate plane turbulent jets as was suggested by the extensive RANS simulations conducted by \textit{Khayrullina et al. }\cite{Khayrullina2019}.\\

\subsection{\textbf{Computational setup and boundary conditions}}
\vspace{0.25cm}
\noindent As described in the preceding section, we study the positively buoyant AC using RANS-based two-dimensional numerical simulations. Towards this, the computational box employed in this study, the size of which is $L =$ 2.1 m, $H =$ 0.35 m, along with the corresponding boundary conditions, is shown in Fig. \ref{fig:domain}.
For the present simulations, the space under consideration (indoor) that requires sealing spans from the left wall ($x$ = 0) to the domain center ($x$ = 1.05 m), and is initially filled with a cold fluid ($T$ = 25$^\circ$C). An air curtain is provided at the top, just inside the building space, that blows air at a temperature lesser or equal to that of the indoor fluid. In this work, we report simulations with the jet temperatures ranging between 15$^\circ$C and 25$^\circ$C. Outside the doorway, a warm fluid ($T$ = 45 $^\circ$C) fills in the rest of the computational domain thus resulting in a reduced gravity, $g^\prime = g \frac{\Delta \rho}{\rho_{av}} \approx $6.7 cm/s$^2$. Here, $\rho_{av}$ is taken as the average density of the indoor and outdoor fluid densities. This setup replicates the temperature levels prevalent in the Indian subcontinent in summer situations and thus seeks optimization of the thermal comfort of building occupants using air curtains. We use ten grid points across the nozzle width at the AC exit based on the recommendation of \cite{krajewski2015air} and a grid independence study (not shown) where the jet statistics were evaluated and compared with the data available in the literature. In order to resolve the spatial instabilities in the air curtain jet and the sharp density gradient at the doorway location, the central region of the domain is locally refined symmetrically (refer to the hatched region in Fig. \ref{fig:domain}). This local refinement allows accurate estimation of the volumetric exchange that primarily initiates at the center of the domain as a result of initial horizontal stratification. At the jet exit, which has a nozzle width of $b_0 =$ 5 mm, a top-hat velocity profile is supplied, and the supply temperature is systematically varied to study its effect on the temperature distribution in the indoor space and the corresponding sealing effectiveness. The resulting doorway height to nozzle-width ratio, $\frac{H}{b}$, is fixed at 84 in our simulations which is similar to that of real-scale air curtain installations \cite{Costa2006}. The side walls of the computational domain are assigned as outflows to simulate a large ambient situation. We also conducted some simulations using a `pressure-outlet' boundary condition, and no significant differences were observed in bulk flow estimates. All the other walls of the computational domain are treated as insulated with no slip and no fluid penetration across them.\\
 
\subsection{\textbf{Solver settings}}
\vspace{0.25cm}
\noindent The ANSYS FLUENT package is used to solve the governing equations of fluid flow and heat transfer. We employ a second-order upwind scheme to approximate the spatial derivatives present in the advective and viscous terms of the governing equations and that associated with the turbulence model. To discretize the temporal field, a first-order implicit method is used, and the time step requirement is constrained by the CFL number of 0.8. The coupled algorithm is used to link the pressure and velocity fields in these incompressible simulations. Three different jet temperatures are simulated in the present work: 15$^\circ$C, 20$^\circ$C and 25$^\circ$C that results in the reduced gravity between the indoor and curtain fluid, $g^\prime = g \frac{\Delta \rho_{IC}}{\rho_0}$ of approx. 2 cm/s$^2$, 1 cm/s$^2$ and 0. For each of these cases, eight simulations are conducted at various values of $D_m$ ranging between 0.05 and 1.5. An additional simulation is also conducted without an air curtain installed, i.e., $D_m$ = 0. This scenario is formally known as a lock-exchange flow (LEF) \cite{Agrawal2021} and has been used as a base case for comparison with the air curtain cases. The exchange flow associated with the LEF case is mathematically evaluated as $q_{OD}$, whereas that with the AC installations is estimated as $q$ (refer eq. \ref{eq:effectiveness}). Thus, a total of 25 distinct simulations are reported in this paper.\\

\section{\textbf{RESULTS AND DISCUSSION}}\label{sec:results}
\vspace{0.25cm}
\noindent The results from the present simulations are now discussed. The adopted simulation methodology based on the 2D RANS simulation and the RNG k-$\epsilon$ turbulence model was first validated against the analytical solution and data available in the literature. The validation was performed for the cases of a pure jet (equivalent to an air curtain under isothermal conditions) and that of a neutrally buoyant air curtain where the buoyancy difference between the air curtain and indoor fluid is zero. These validation metrics are discussed in detail in \textit{Agrawal et al.} \cite{agrawal2021towards} and are not presented here. Overall, a good agreement is observed using the adopted computational methodology.

\subsection{\textbf{Effect of deflection modulus}}\label{dmeffect}
\begin{figure}[h!]
\centering
\includegraphics[width=1.0\textwidth]{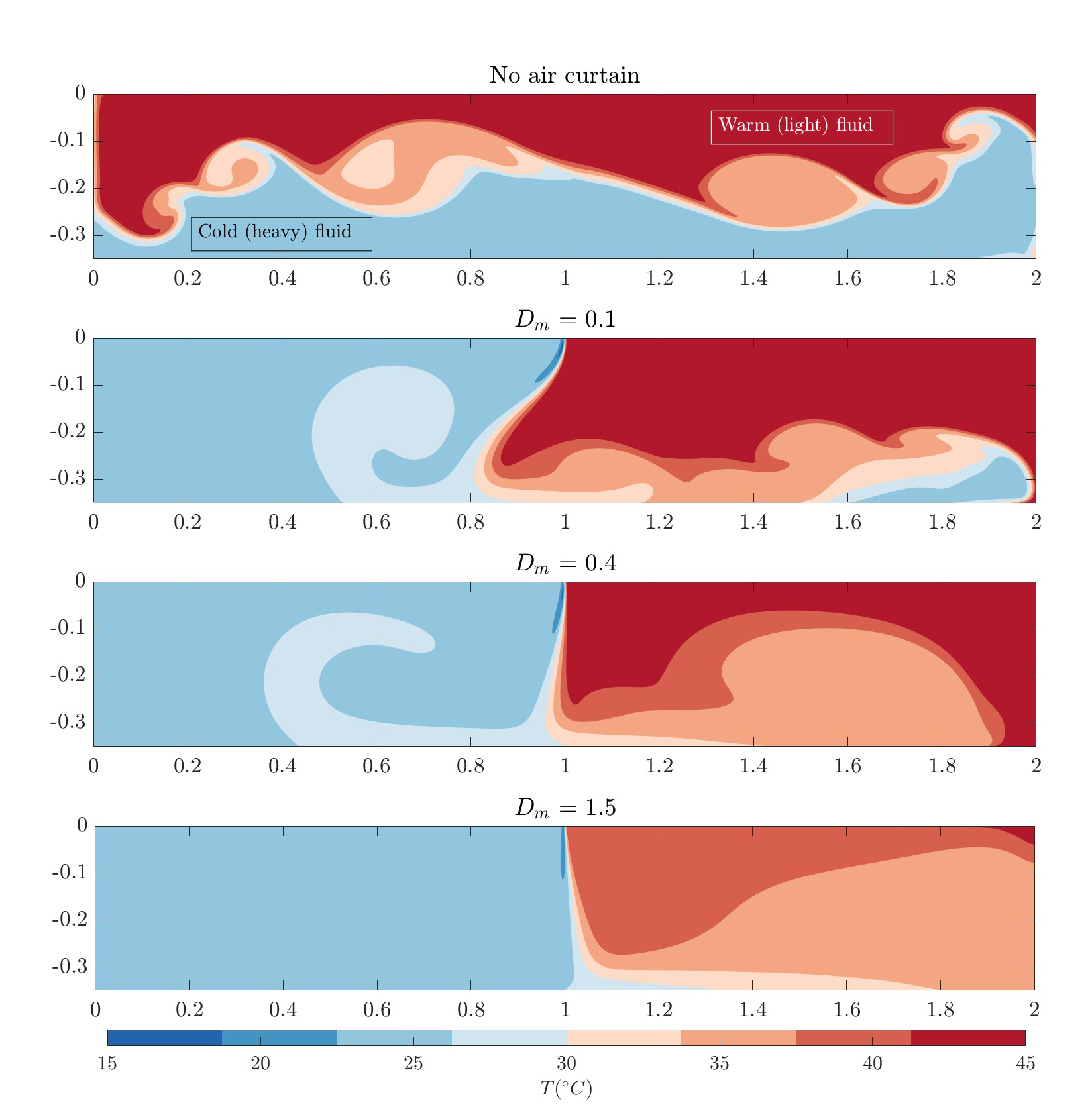}
\caption{Temperature distribution ($^\circ$C) in the domain for different values of $D_m$ at $t^* = t\sqrt{\frac{g^\prime}{H}} =$ 7 and with $T_0 =$ 15$^\circ$C.}
\label{fig:tempcontoursdiffdm}
\end{figure}

\vspace{0.25cm}
\noindent First, we present the temperature distribution in the computational domain for different values of deflection modulus to illustrate the effect of jet `strength' on the aerodynamic sealing in a qualitative way. Figure \ref{fig:tempcontoursdiffdm} shows the temperature contours for the cases of $D_m$ of 0, 0.1, 0.4 and 1.5 when the jet inlet temperature is 15$^\circ$C. The corresponding no AC case, i.e., the LEF scenario, as illustrated in the top panel, allows a free exchange of the two fluids across the doorway. This results in a gradual increment in the room temperature following convection-driven bulk transport, mixing, and thermal diffusion, which is highly undesirable from a thermal comfort perspective. In comparison, when an air curtain is installed, this exchange is significantly suppressed, as shown in the other panels.\\

\noindent When the jet is relatively weaker ($D_m \leq$ 0.1), it deflects towards the dense fluid owing to buoyancy and pressure forces and also allows some buoyancy-driven exfiltration (see the blue patch beyond $x >$ 1.6 $m$ in the second panel). Such AC installations are usually termed unstable or breakthrough as they do not impose a complete sealing across the height of the doorway. Irrespective of the magnitude of positive buoyancy imposed at the jet exit, we observe similar behavior of positively buoyant AC at small values of $D_m$, which is consistent with their neutrally buoyant counterpart. In the present study, we increase the initial jet momentum flux to increase $D_m$ and observe that the jet impinges almost vertically on the floor at higher values of $D_m$ (see the third and fourth panels). This ensures proper sealing across the doorway, and the air curtain installation is deemed \textit{stable}. No exfiltration is observed at these values of deflection modulus, and the primary source of bulk fluid transport appears to stem from the fluid entrainment and turbulent mixing along the edges of the jet and near the impingement region. When the deflection modulus increases further, it can be observed that the fluid is very well mixed as compared to the other cases. The quantitative implications of the deflection modulus on sealing effectiveness are discussed in the next section.

\subsection{\textbf{Sealing effectiveness}}\label{positivecurtains}
\vspace{0.25cm}
\noindent Based on eq. \ref{eq:effectiveness}, the performance of an air curtain is obtained using the fraction of exchange flow that it suppresses as compared to the case of a doorway with no air curtain installed. In the absence of an AC, the volumetric exchange can be theoretically estimated using the orifice equation as:
\begin{equation}
    q_{OD} = \frac{1}{3} C_d A \sqrt{g^\prime H}
\end{equation}
Here, $C_d$ represents the coefficient of discharge, and $A$ is the cross-sectional area available at the exchange site and is taken as the doorway height for the present 2D computations. The exchange flow, $q_{OD}$, obtained from our simulations for the case of $Q =$ 0, results in a $C_d$ of 0.56, which agrees very well with the empirical estimate of 0.6 for a rectangular orifice. To estimate the exchange flow in the presence of an air curtain, we use the mass conservation in the indoor space \cite{Jha2021b}:
\begin{equation}
    q = \frac{V_0}{t} \left (\frac{\rho_d - \overline{\rho}}{\rho_d - \rho_l} \right) + {\beta Q} \left (\frac{\rho_0 - \overline{\rho}}{\rho_d - \rho_l} \right)
\end{equation}
Here, $V_0$ is the constant volume of the indoor space, and $Q$ denotes the volumetric flow rate (in m$^2$/s) supplied by the AC per unit width. $\beta$ represents the spillage fraction of the AC jet fluid inside the room after impingement on the floor. We take $\beta$ as 0.5 following the discussion in \cite{Jha2021b}. Lastly, $\overline{\rho}$ is the mean fluid density inside the room at any time instant $t$. \\
\begin{figure}[h!]
    \centering
    \begin{subfigure}{0.95\linewidth}
    \centering
    \includegraphics[width=\textwidth]{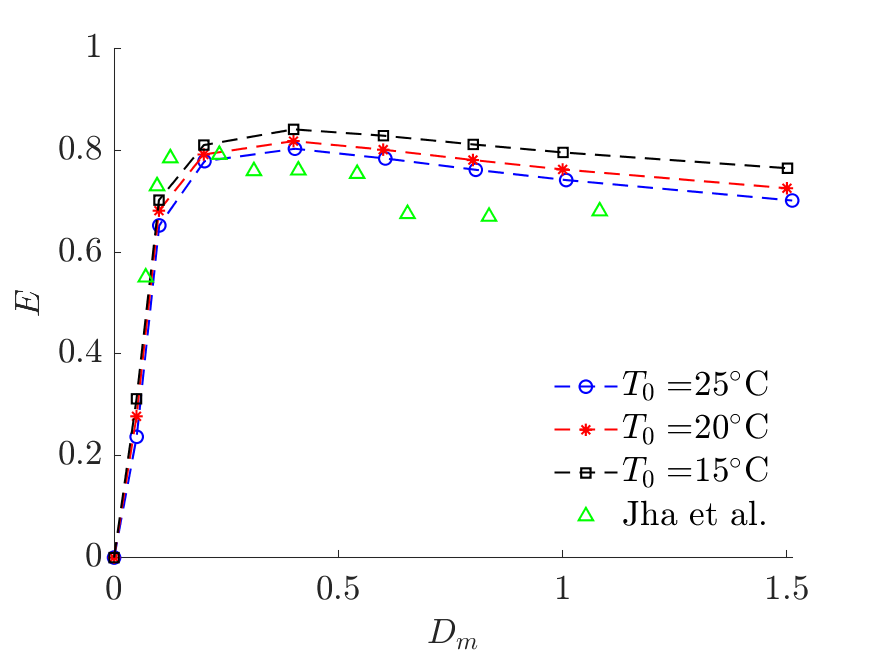}
    \caption{Effectiveness as a function of deflection modulus for all the present simulations.}
  \label{fig:effectiveness_all}
    \end{subfigure}
    \begin{subfigure}{0.95\linewidth}
    \centering
    \includegraphics[width=\textwidth]{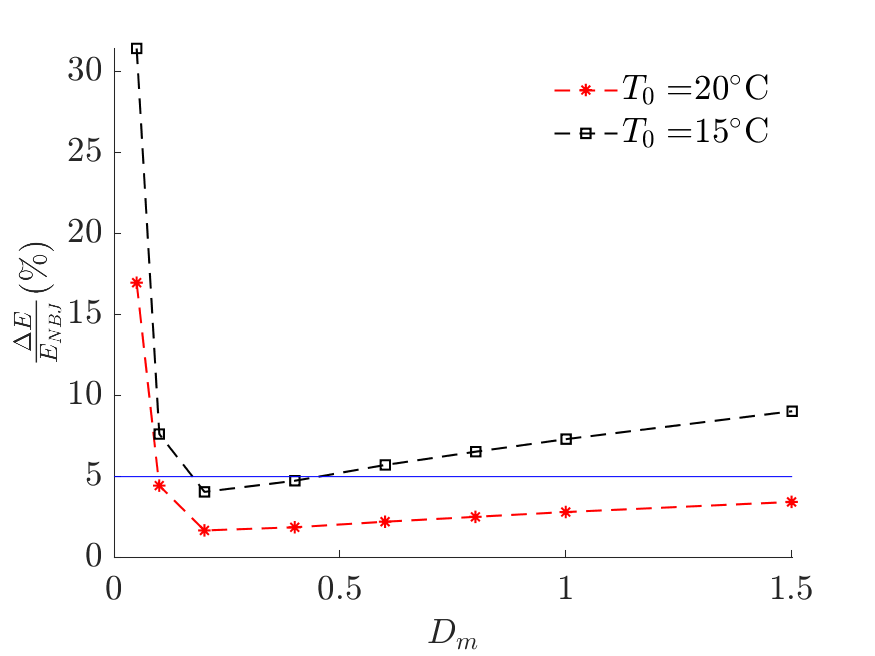}
    \caption{Percentage increase in effectiveness as compared to the neutrally buoyant AC.}
  \label{fig:deltaE}
    \end{subfigure}
    \caption{Sealing effectiveness of the positively buoyant air curtains.}
    \label{fig:effectivenessplots}
\end{figure}

\noindent The sealing effectiveness estimates for all the simulations reported in the present study are shown in figure \ref{fig:effectiveness_all} and also compared with the experimental estimates of \textit{Jha et al.} \cite{Jha2021b} who measured the effectiveness of neutrally buoyant AC using bulk density measurement techniques. First, it can be seen that a good agreement exists between the present neutrally buoyant AC and those of \textit{Jha et al.}. While the RANS simulations over predict the quantitative values by a certain amount, especially at higher deflection modulus, the overall behavior with varying $D_m$ is captured well. At low values of deflection modulus, owing to jet deflection, the effectiveness is rather small for the breakthrough case. Once the $D_m$ increases, the sealing becomes more prominent. We obtain the peak effectiveness in the range of $D_m \approx$ 0.2 - 0.4, which is similar to that of experimental results. Upon increasing the deflection modulus further, there is a gradual decrease in $E$, which is usually attributed to more mixing in the computational domain due to turbulence. In comparison to the neutrally buoyant AC, upon supplying the air curtain fluid with positive buoyancy, we observe an increase in sealing effectiveness. This is because the additional buoyancy in the AC jet assists in maintaining the indoor fluid density (temperature), and as a result, the corresponding density difference, $\rho_d - \overline{\rho}$, decreases, resulting in larger effectiveness. In extreme cases, it is also possible that the added buoyancy in the air curtain can nullify the effects of fluid transport from the outdoor space. However, that possibility was not explored in the present study and could be considered in future studies.\\

\noindent As illustrated in figure \ref{fig:deltaE}, the percentage increase in the sealing effectiveness, as compared to its neutrally buoyant counterpart, is a weak function of $D_m$ when the air curtain is stable ($D_m >$ 0.1). For smaller values of $D_m$, the relatively higher $\Delta E$ is due to the deflection of positively buoyant air curtain inside the room, which increases the mean indoor fluid density. For the cases of $T_0$ being 15$^\circ$C and 20$^\circ$C, we calculate the maximum relative increment in the effectiveness of stable AC to be of the order of approx 10\%. Whereas the maximum sealing effectiveness is found to be 5\% larger than its corresponding value for a neutrally buoyant air curtain. This maximum is achieved at the deflection modulus of approximately 0.4.\\

\subsection{\textbf{Fractional spillage of the AC jet}}\label{beta_jet}
\vspace{0.25cm}
\noindent In this section, we present estimates of $\beta$ for the case of positively buoyant AC for varying deflection modulus. To compute this, we introduce a passive scalar in the air curtain jet and compute the mean spatial concentration of this scalar in the computational domain. In isothermal case, i,e. $\rho_l = \rho_d = \rho_0$, the jet divides itself symmetrically across the doorway i.e. $\beta =$ 0.5, if the outlets are symmetrical. However, it is known that the transverse buoyancy affects the longitudinal path traveled by jet, especially at lower values of $D_m$ (see the second panel in figure \ref{fig:tempcontoursdiffdm}). Therefore, the fractional spillage, $\beta$, might differ from the isothermal value of 0.5 if the effect of transverse buoyancy is substantial. The estimates of $\beta$ are inherently difficult to obtain through experiments due to the number of species involved in such measurements. Generally, either salt or heat is used to create density differences required for the initial horizontal stratification. If detailed quantitative information is sought, then other scalars, for example, a fluorescent dye to measure the scalar field, are also used. However, direct measurements of a passive scalar to compute $\beta$ have not been reported to the best of our knowledge.\\
\begin{figure}[h!]
    \centering
    \includegraphics[width = \textwidth]{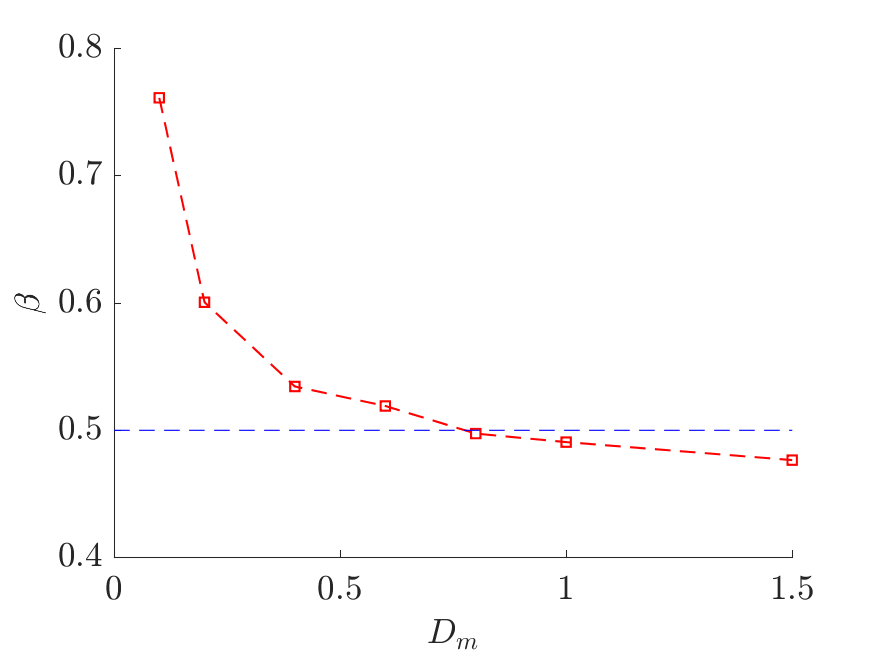}
    \caption{Spillage fraction, $\beta$, for different values of $D_m$ with $T_0 =$ 15$^\circ$C.}
    \label{fig:beta}
\end{figure}

\noindent Figure \ref{fig:beta} shows the variation of the spillage fraction, $\beta$, for various deflection modulus for the case of $T_0 =$ 15$^\circ$C. The unusually high value of $\beta$ at low $D_m$ of 0.1 is due to the jet deflection (refer to the figure \ref{fig:tempcontoursdiffdm}). However, once the air curtain is stable, the nominally assumed value of $\beta =$ 0.5 appears to be a reasonable estimate. In the range of deflection modulus between 0.4 and 1.5, this does not introduce an error of more than 5\% in the estimates of sealing effectiveness. An interesting observation from figure \ref{fig:beta} is that $\beta$ reduces slightly with an increase in $D_m$. This might be attributed to the different density fluids in the computational domain in which the air curtain jet penetrates. Consider the indoor space, for example, where the effective density difference for the incoming jet $\rho_0 - \rho_d$, is smaller than its ambient counterpart, $\rho_0 - \rho_l$. This results in a larger penetrating force in the outside space and, thus, a smaller $\beta$.\\

\section{\textbf{CONCLUSIONS}}\label{sec4}
\vspace{0.25cm}
\noindent This paper reported 2D simulations of positively buoyant air curtain flows using the Reynolds-averaged Navier-Stokes (RANS) methodology. The computations employed the RNG k-$\epsilon$ model for turbulence closure and were performed in ANSYS FLUENT. The sealing effectiveness of air curtains was estimated using these simulations for a range of deflection modulus between 0.05 and 1.5 and for three different jet temperature levels. It was shown that the sealing effectiveness increased as the extent of positive buoyancy increased. In comparison to the neutrally buoyant case, the peak effectiveness of the positively buoyant AC was found to be 5\% larger, whereas the maximum increment was computed to be approximately 10\%. The jet spillage fraction was also estimated by introducing a passive scalar in the air curtain jet. It was observed that the isothermal value of 0.5 does not introduce errors of more than 5\% in the computation of sealing effectiveness when the deflection modulus is larger than approximately 0.4.\\

\vspace{0.5cm}
 \noindent
\textbf{NOMENCLATURE}\\

\begin{tabular}{ccc}
$Re$ & Reynolds number & -- \\
$\beta$ & Spillage fraction & --- \\
$t^*$ & Dimensionless time & -- \\
$D_m$ & Deflection modulus & -- \\
$E$ & Sealing effectiveness & -- \\
$C_d$ & Coefficient of discharge & -- \\
$t$ & Time & [s] \\
$A$ & Cross-sectional area & [m$^2$] \\
$b_0$ & Nozzle width at exit & [m] \\
$T_0$ & Exit jet temperature & [K] \\
$H$ & Doorway height & [m] \\
$u_0$ & Nozzle velocity at exit & [m/s] \\
$u_i$ & $i^{th}$ velocity component & [m/s] \\
$g^\prime$ & Reduced gravity & [m/s$^2$] \\
$q$ & Exchange flow rate & [m$^3$/s] \\
$q_{OD}$ & Exchange flow without air curtain & [m$^3$/s] \\
$p$ & Pressure & [Pa] \\
$T$ & Temperature & [K] \\
$\rho$ & Density & [kg/$m^3$] \\
$\mu$ & Dynamic viscosity & [Pa$\cdot$s] \\
\end{tabular}

\bibliographystyle{ieeetr}
\bibliography{all_cite}

\begin{thebibliography}{10}

\bibitem{Goncalves2012}
J.~C. Gon{\c{c}}alves, J.~J. Costa, A.~R. Figueiredo, and A.~M. Lopes, ``{CFD
  modelling of aerodynamic sealing by vertical and horizontal air curtains},''
  {\em Energy and Buildings}, vol.~52, pp.~153--160, 2012.

\bibitem{Hayes1969Design}
F.~C. Hayes and W.~F. Stoecker, ``{Design Data For Air Curtains},'' {\em ASHRAE
  Transactions}, pp.~168--180, 1969.

\bibitem{Costa2006}
J.~J. Costa, L.~A. Oliveira, and M.~C. Silva, ``{Energy savings by aerodynamic
  sealing with a downward-blowing plane air curtain-A numerical approach},''
  {\em Energy and Buildings}, vol.~38, no.~10, pp.~1182--1193, 2006.

\bibitem{Foster2007}
A.~M. Foster, M.~J. Swain, R.~Barrett, P.~D'Agaro, L.~P. Ketteringham, and
  S.~J. James, ``{Three-dimensional effects of an air curtain used to restrict
  cold room infiltration},'' {\em Applied Mathematical Modelling}, vol.~31,
  no.~6, pp.~1109--1123, 2007.

\bibitem{Frank2014}
D.~Frank and P.~F. Linden, ``{The effectiveness of an air curtain in the
  doorway of a ventilated building},'' {\em Journal of Fluid Mechanics},
  vol.~756, pp.~130--164, 2014.

\bibitem{Jha2021b}
N.~K. Jha, D.~Frank, and P.~F. Linden, ``{Contaminant transport by human
  passage through an air curtain separating two sections of a corridor: Part I
  – Uniform ambient temperature},'' {\em Energy and Buildings}, vol.~236,
  p.~110818, 2021.

\bibitem{Jha2020b}
N.~K. Jha, D.~Frank, L.~Darracq, and P.~F. Linden, ``{Contaminant transport by
  human passage through an air curtain separating two sections of a corridor:
  Part II - two zones at different temperatures},'' {\em Energy and Buildings},
  vol.~236, p.~110728, 2021.

\bibitem{agrawal2021towards}
T.~Agrawal, N.~K. Jha, and V.~K. Chalamalla, ``Towards understanding air
  curtain flows using rans based numerical simulations,'' 2021.

\bibitem{howell1980optimum}
R.~H. Howell and M.~Shibata, ``{Optimum Heat Transfer Through Turbulent
  Recirculated Plane Air Curtains},'' {\em ASHRAE Transactions}, vol.~2567,
  pp.~188--200, 1980.

\bibitem{frank2015effects}
D.~Frank and P.~Linden, ``The effects of an opposing buoyancy force on the
  performance of an air curtain in the doorway of a building,'' {\em Energy and
  Buildings}, vol.~96, pp.~20--29, 2015.

\bibitem{yakhot1992development}
V.~Yakhot, S.~Orszag, S.~Thangam, T.~Gatski, and C.~Speziale, ``Development of
  turbulence models for shear flows by a double expansion technique,'' {\em
  Physics of Fluids A: Fluid Dynamics}, vol.~4, no.~7, pp.~1510--1520, 1992.

\bibitem{Khayrullina2019}
A.~Khayrullina, T.~van Hooff, B.~Blocken, and G.~van Heijst, ``{Validation of
  steady RANS modelling of isothermal plane turbulent impinging jets at
  moderate Reynolds numbers},'' {\em European Journal of Mechanics, B/Fluids},
  vol.~75, pp.~228--243, 2019.

\bibitem{krajewski2015air}
G.~Krajewski and W.~Wegrzynski, ``Air curtain as a barrier for smoke in case of
  fire: Numerical modelling,'' {\em Bulletin of the Polish Academy of Sciences:
  Technical Sciences}, pp.~145--153, 2015.

\bibitem{Agrawal2021}
T.~Agrawal, B.~Ramesh, S.~J. Zimmerman, J.~Philip, and J.~C. Klewicki,
  ``{Probing the high mixing efficiency events in a lock-exchange flow through
  simultaneous velocity and temperature measurements},'' {\em Physics of
  Fluids}, vol.~016605, 2021.

\end{thebibliography}

\end{document}